\documentclass[twocolumn]{aastex63}
\usepackage{graphicx}
\usepackage{amsmath}   
\usepackage{amssymb}

\usepackage{graphicx}  
\usepackage{amsmath}   
\usepackage{amssymb}  
\usepackage{color}

\usepackage{natbib}

\usepackage[active]{srcltx}

\def\eq#1{{Eq.~(\ref{#1})}}

\begin{document}

\title{Alleviating the need for exponential evolution of JWST galaxies in 10$^{10} M_{\odot}$ haloes at $z > 10$ by a modified $\Lambda$CDM power spectrum}

\author{Hamsa Padmanabhan}
\affiliation{ D\'epartement de Physique Th\'eorique, Universit\'e de Gen\`eve \\
24 quai Ernest-Ansermet, CH 1211 Gen\`eve 4, Switzerland\\
}

\author{Abraham Loeb}
\affiliation{Astronomy department, Harvard University \\
60 Garden Street, Cambridge, MA 02138, USA}

\email{hamsa.padmanabhan@unige.ch, aloeb@cfa.harvard.edu}

  \begin{abstract}
We infer the evolution of the UV luminosities of galaxies in haloes of masses $10^{10} - 10^{11} M_{\odot}$ in the redshift range of  $z \sim 9-16$ from the recent JWST data. Within the standard $\Lambda$CDM cosmological model,
it is found that the average luminosities in this halo mass range show an exponential evolution with redshift, in excess of that expected from astrophysical considerations including the evolution of UV luminosity from Population III galaxies. We find that an enhancement of power on scales $k \sim 1$ Mpc$^{-1}$, as captured by a cosmological transfer function modified  from the $\Lambda$CDM form, is able to alleviate this effect and allow for a non-evolving UV luminosity as a function of redshift at $z > 10$, consistently with the corresponding findings for lower redshifts. We discuss the possible astrophysical and cosmological reasons for such an enhancement.
 \end{abstract}

\keywords{cosmology: theory -- dark ages, reionization, first stars --   early Universe -- galaxies: high-redshift}

\section{Introduction}
The early-release photometric surveys from James Webb Space Telescope (JWST) have revealed unprecedented insights into the nature of the bright galaxies out to high redshifts, $z \sim 16$ \citep{atek2023, naidu2022, finkelstein2023, castellano2022, treu2022, mcleod2023,  harikane2023, papovich2023, donnan2023, yan2023, adams2023a, castellano2023b}, with spectroscopic confirmation of many objects now available \citep{robertson2023, curtislake2023, arrabalharo2023}. Most semi-analytical and hydrodynamical simulations of galaxy formation are consistent with the observed galaxy properties out to $z \sim 9-10$, with deviations required to explain the observed abundances and luminosities  at higher redshifts, such as a more aggressive star-formation rate evolution or feedback-free starbursts \citep[e.g.,][]{kannan2022,qin2023, labbe2023, dekel2023, lovell2023, adams2023, shen2023, trinca2023}. Possible modifications to the $\Lambda$CDM cosmology \citep[e.g.,][]{liu2022, biagetti2023, parashari2023, jiao2023,hutsi2023} have also been proposed with a view to reduce the tension between observations and theoretical predictions at $z > 10$.
 
The UV luminosity functions (UV LFs) of galaxies (measured at the  rest-frame wavelength of 1500 \AA) has been extensively used to probe the cosmological evolution of dark matter haloes out to $z \sim 8$ \citep[e.g.,][]{mashian2015, rudakovskyi2021, sabti2021, ferrara2023}.  \citet{mashian2015} employed abundance matching techniques on the UV LFs at $z < 8$ and found that the average UV luminosity (and equivalent star-formation rate) remains roughly constant at any given dark matter halo mass, allowing for the development of an average star-formation rate - halo mass relation over $4 < z < 8$. Here, we derive the corresponding relation out to $z \sim 16$ using the luminosity functions calibrated from JWST data by \citet{harikane2023} and \citet{donnan2023}. We find that the average UV luminosity of galaxies in haloes of masses $10^{10} - 10^{11} M_{\odot}$ within the $\Lambda$CDM framework shows a strong evolution as a function of redshift,  above the levels expected from the astrophysical evolution of the star formation rate and the fraction of Population III stars. We show that a power spectrum modified from the  $\Lambda$CDM form, incorporating an enhancement of power on scales of $\sim 1 \ {\rm Mpc}^{-1}$ is able to alleviate this sharp exponential rise and reproduce the non-evolving nature of the UV luminosity - halo mass relation at all redshifts.

The paper is organized as follows. In Sec. \ref{sec:abmatch}, we overview the data used to calibrate the average UV luminosity - halo mass relation over $z \sim 9-16$ and present the inferred evolution of the luminosity as a function of redshift using the abundance matching technique. In Sec. \ref{sec:astro}, we discuss the challenges associated with various astrophysical explanations for the observed evolution, including the expected growth of the Population III stellar fraction in haloes. We then introduce (Sec. \ref{sec:modtrans}) an enhancement in the power on scales of $\sim 1 \ {\rm Mpc}$, achieved by a modification to the $\Lambda$CDM transfer function, which is found to  lead to a non-evolving form of the UV luminosity in line with the corresponding behaviour at $4 < z < 8$. We discuss the possible implications of these results in Sec. \ref{sec:discussion}.

\section{JWST observations and abundance matching}
\label{sec:abmatch}
We use the photometric UV LFs calibrated over $z \sim 9-16$ from the JWST early-release NIRCam imaging data \citep{donnan2023, harikane2023} covering the SMACS0723 target, the GLASS JWST Early Release Science \citep[GLASS;][]{treu2022}, and the Cosmic Evolution Early Release Science \citep[CEERS;][]{finkelstein2023} fields. The data analyses provide fitting forms for the number density (including uncertainties) of sources per absolute magnitude at the redshifts 9, 10.5, 12, 13.25 and 16.\footnote{ It is to be noted that the photometric JWST detections  at $z > 10$ are very sparsely sampled, although the statistics permit the development of luminosity functions. For example, \citet{harikane2023} find no candidates between $z \sim 12.28$ and $z \sim 16.25$, and there are a total of about 23 systems found over $z \sim 9-16$.}
Abundance matching of the UV LFs to the Sheth-Tormen \citep{sheth2002} form\footnote{The abundance matching procedure assumes the luminosity of the galaxies to be a monotonic function of their host dark matter halo mass. This is justifiable in the present context since the halo substructure does not make a significant contribution \citep{mashian2015} and the clustering strength of galaxies increases with their UV luminosity similarly to the increase of halo clustering with host halo mass \citep{lee2009}. Since the galaxies have very young ages, they are expected to shine for a significant fraction of 100 Myr -- the approximate age of the Universe at these epochs -- and thus not expected to undergo several bursts of star formation.} of the comoving number of haloes per logarithm of halo mass, $dn/d \log_{10} M_h$:
\begin{eqnarray}
 &&  \int_{M_h (L_{\rm UV})}^{\infty} \frac{dn}{ d \log_{10} M'_h} \ d \log_{10} M'_h \nonumber \\
  &=& \int_{L_{\rm UV}}^{\infty} \phi(L_{\rm UV}) \ d \log_{10} L_{\rm UV}
  \label{abmatchuvlf}
\end{eqnarray}
leads to the inferred UV luminosity $L_{\rm UV}$ as  a function of host halo mass $M_h$ at any given redshift. Here, $\phi(L_{\rm UV}$)  is the comoving number density of galaxies per logarithm of UV luminosity $L_{\rm UV}$.

\begin{figure}
\centering
\includegraphics[width = \columnwidth]{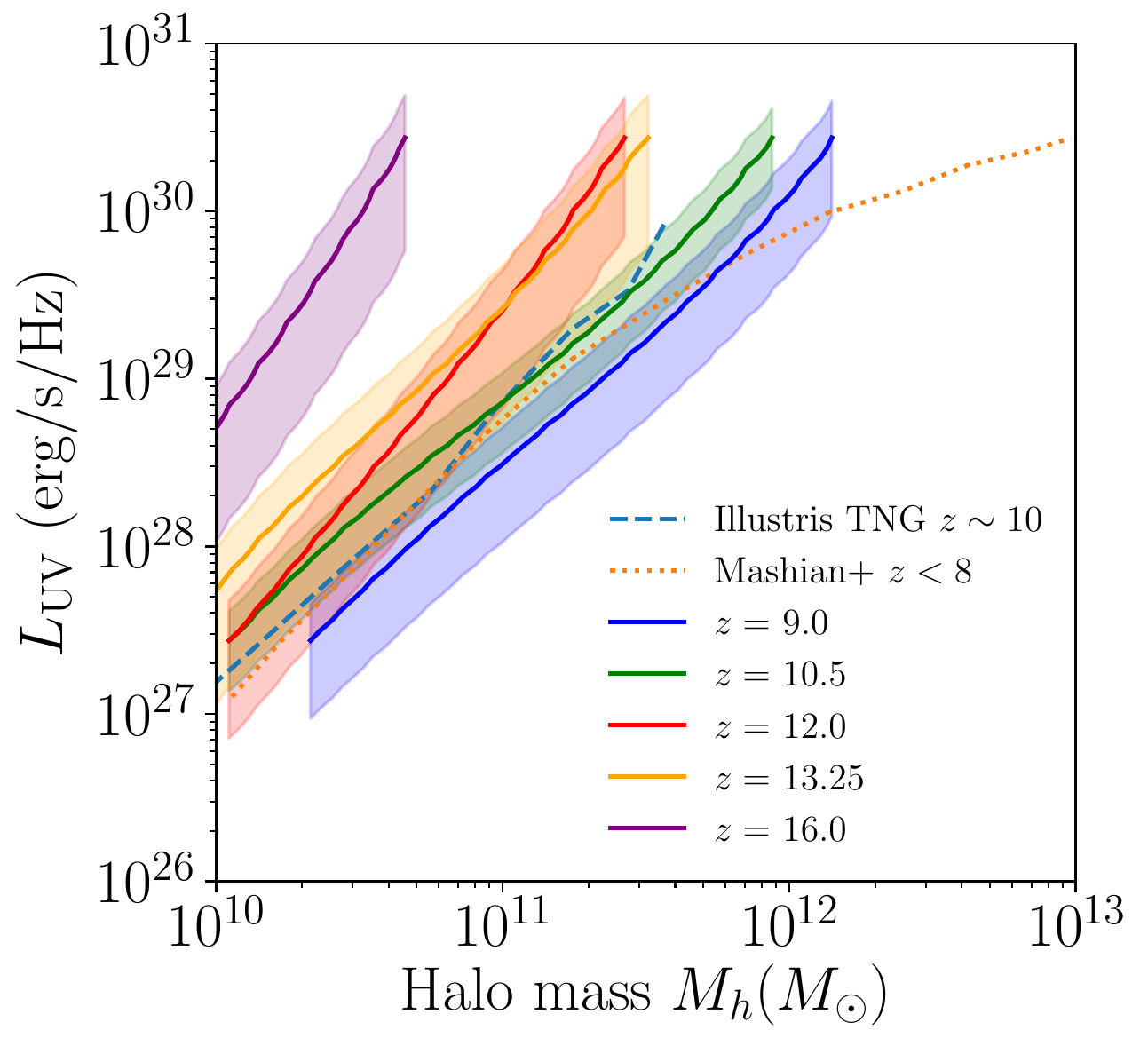}
\caption{The average UV luminosity as a function of redshift and halo mass from the JWST observations { (solid lines) with the shaded bands denoting the uncertainties inferred from the UV LFs.} Also plotted are the predicted luminosity - halo mass relations from the simulations of \citet[][dashed blue line]{kannan2022} and the empirical findings at lower redshift \citep[][orange dotted line]{mashian2015}.}
\label{fig:speclum}
\end{figure}
The inferred UV luminosity - host halo mass relation obtained from the abundance matching procedure is plotted in Fig. \ref{fig:speclum}, which shows that the haloes in the mass range $10^{10}$ - $10^{11} \ M_{\odot}$  exhibit a sharp rise in their average UV luminosity  at $z > 10$, in contrast to the findings at $z \sim 4-8$. This is also seen from the comparison to the empirical findings of \citet{mashian2015} at lower redshifts, $z < 8$ where an average UV luminosity - halo mass relation could be derived (orange dotted line) and the IllustrisTNG simulations  at $z \sim 10$ \citep[][dashed blue line]{kannan2022}.  These results are found to be broadly consistent with theoretical expectations up to $z \sim 10$, requiring deviations  above this redshift \citep{kannan2022}.

The UV luminosity over $z \gtrsim 10$ at a fixed halo mass (between $10^{10}$ and $10^{11}$ $M_{\odot}$) is plotted in Fig. \ref{fig:uvevol} with the uncertainty band derived from the observational uncertainty in the LF. The average relation is found to evolve exponentially with redshift. 
The best-fitting functional form is plotted in Fig. \ref{fig:uvevol} as the dotted purple line, which goes as
\begin{equation}
\log_{10} (\langle L_{\rm UV} \rangle/ {\rm ergs/s/Hz}) = A + Bz
\label{uvfit}
\end{equation}
at $z > 10$, and
\begin{equation}
\log_{10} (\langle L_{\rm UV} \rangle/ \rm ergs/s/Hz) = {\rm const}
\end{equation}
at $z < 10$.
The values of the best-fitting parameters are $A = 25.7 \pm 0.23$, $B = 0.25 \pm 0.02$ and const $= 28.1$.
The behaviour at $z > 10$ is very different to that at lower-$z$ where the UV luminosity (or equivalent star formation rate) was found to be a fairly non-evolving function of halo mass \citep[e.g.][]{mashian2015}.

\section{Astrophysical explanations}
\label{sec:astro}
The sharp exponential rise in the luminosity for halos in the above mass slice at $z>10$ could be the result of: (i)  the presence of a Population III (Pop III) Initial Mass Function (IMF), as suggested for some systems like GN-z11 \citep[e.g.,][]{maiolino2023, hploeb2022} or lower-metallicity stars that produce a higher fraction of UV-ionizing photons \citep[e.g.,][]{trinca2023}, (ii) a much higher escape fraction { implied by gas being less likely to stay in the haloes}, and/or (iii) a much higher conversion efficiency of gas to stars.  The escape fraction of UV photons at lower redshifts is in the range 5-100\%, and the star formation efficiency is about 10\% \citep{loeb2013, khaire2016}. The number of ionizing (UV) photons per baryon in stars is $\sim 5000$ for a standard IMF, and this number grows to $\sim 10^5$ for a top heavy IMF. This shows that the IMF can change the UV luminosity by a factor of 20 and must be supplemented by the other factors.

\begin{figure}
\centering
\includegraphics[width = \columnwidth]{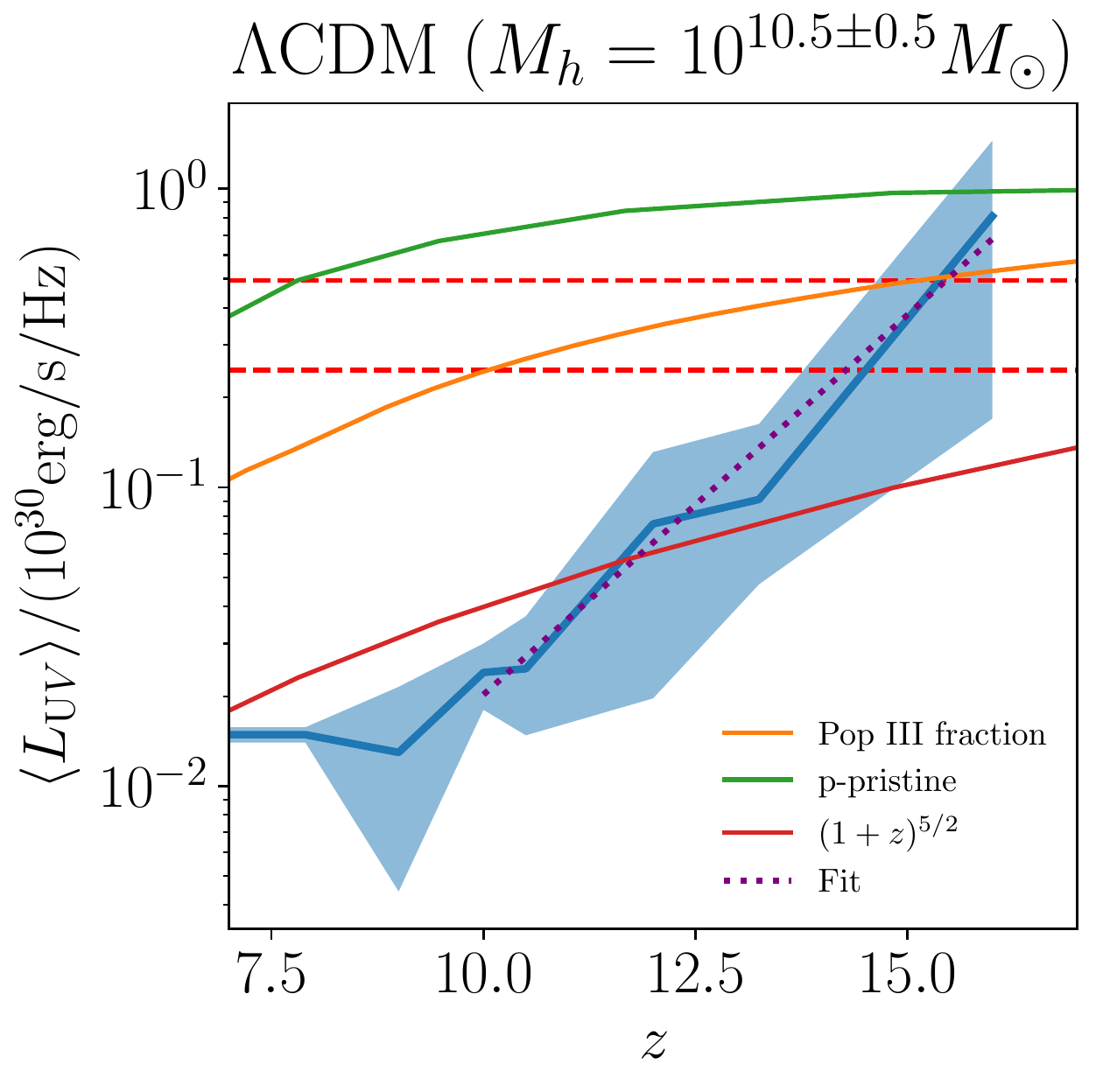}
\caption{The average UV luminosity for $10^{10} - 10^{11} M_{\odot}$ haloes as a function of redshift derived from abundance matching (solid blue line), with its associated uncertainty (shaded blue band). The best-fitting evolution of the average relation (Eq. \ref{uvfit})  
is indicated by the purple dotted line. Overplotted are the expected fractions of Pop III stars (averaged over all haloes) and that of pristine gas (with an arbitrary normalization) from the results of \citet{furlanetto2005, sun2021} and the gas accretion rate, proportional to $(1+z)^{5/2}$ \citep{furlanetto2022}. The red dashed lines indicate the range of increase in UV luminosity from a top-heavy IMF consisting of $300 - 1000 \ M_{\odot}$ stars \citep{bromm2001}.}
\label{fig:uvevol}
\end{figure}

\begin{figure}
\centering
\includegraphics[width = \columnwidth]{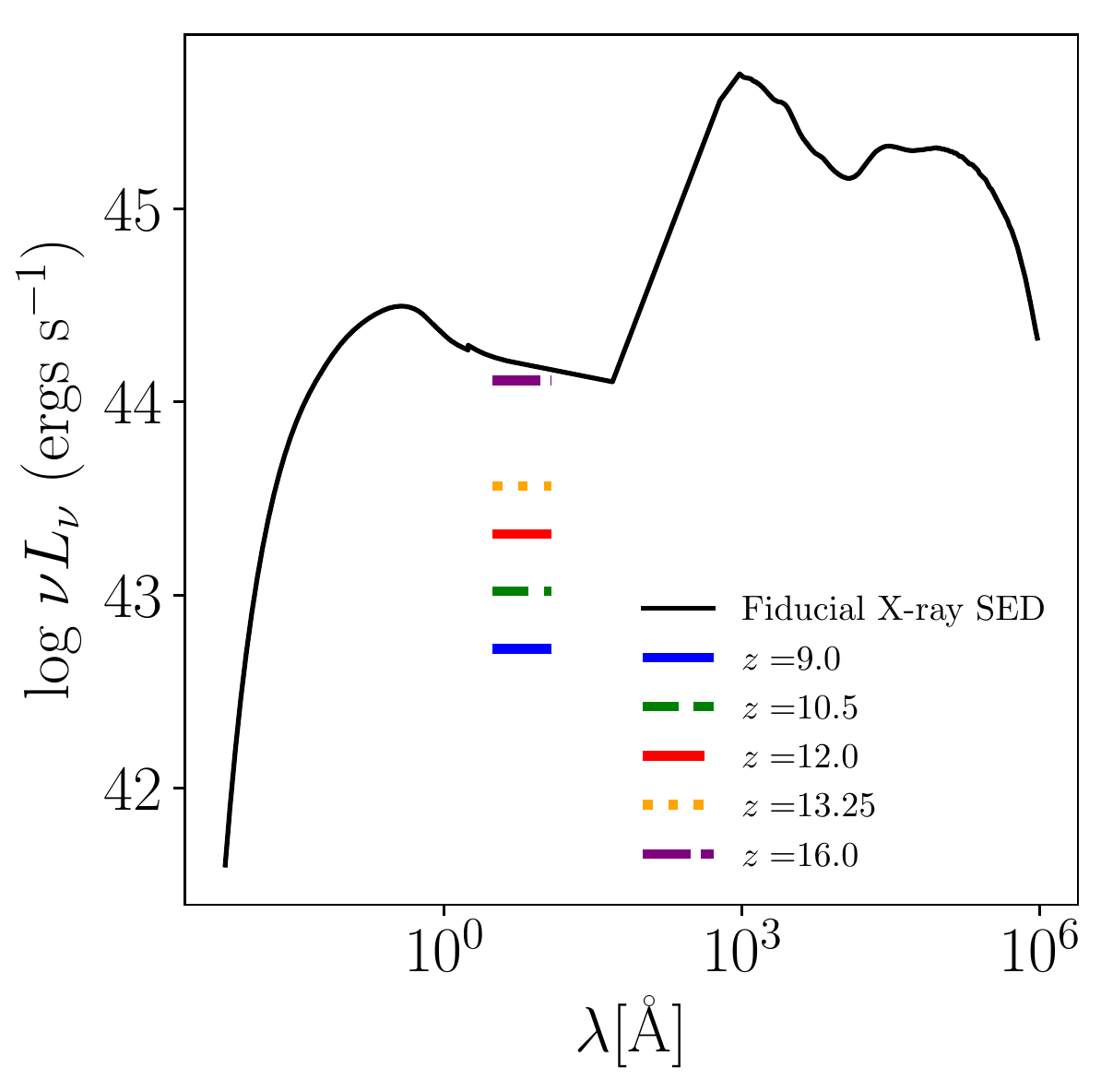}
\caption{Spectral energy density of AGN in the 0.5-7 keV band corresponding to the observed average UV luminosity at various redshifts, compared to a fiducial AGN spectrum at $z \sim 6$ normalized to $\log_{10} (\nu L_\nu/{\rm ergs \ s^{-1}})$ (4500 \AA) = 45.5 \citep{shen2020}.}
\label{fig:sedjwst}
\end{figure} 
{ Photons from galaxies must escape absorption by their surrounding gas and dust in order to contribute to the observed UV luminosity \citep{loeb2013}. At high redshifts, gas is less likely to stay in haloes as its characteristic speed follows the halo circular velocity $v_c \propto M_h^{1/3} (1+z)^{1/2}$ which increases with redshift for a fixed halo mass $M_h$. While the exact behaviour of $f_{\rm esc}$ shows a large variation among galaxies, observations are consistent  with only a slowly rising trend of $f_{\rm esc}$ with redshift, flattening in the pre-reionization era  \citep[][]{khaire2016}.} The efficiency of conversion of gas to stars may also increase from the lower redshift value of $f_* \sim 0.1$, however, a maximum such increase by a factor of 10 is too low to explain the observed rise in luminosity at the highest redshifts.

Various theoretical predictions for the evolution of Pop III stars with redshift have been considered in the literature \citep[e.g., ][]{sun2021, venditti2023,latif2022}. At $z \sim 11-15$, a factor of 10-20 in the UV luminosity could result from a top-heavy IMF that scales the luminosity \citep{bromm2001} as shown by the red dashed lines in Fig. \ref{fig:uvevol} for Pop III stars of masses 300 - 1000 $M_{\odot}$. This is consistent with the expectations of \citet{yung2023} which find a boost of about 4 to an order of magnitude more UV photons are warranted to explain the observed UV luminosities over a range of halo masses. 

However, when this IMF is convolved  with the expected \textit{fraction} of Pop III IMF stars at a given halo mass as a function of redshift, considering that its effect decreases the normalization of the UV luminosity - star formation rate relation to about a third of its Salpeter value \citep{bromm2001, harikane2023b, harikane2023}, the rise of the UV luminosity with redshift becomes connected to that of the Pop III fraction.
For a fixed accretion rate of pristine gas into dark matter haloes, this fraction is expected to go down exponentially following the abundance of haloes at fixed halo mass \citep{sun2021, parsons2022}. A more aggressive rise of the accretion rate with increasing redshift is needed to salvage the match to the observed evolution.
\begin{figure}
\centering
\includegraphics[width = 0.8\columnwidth]{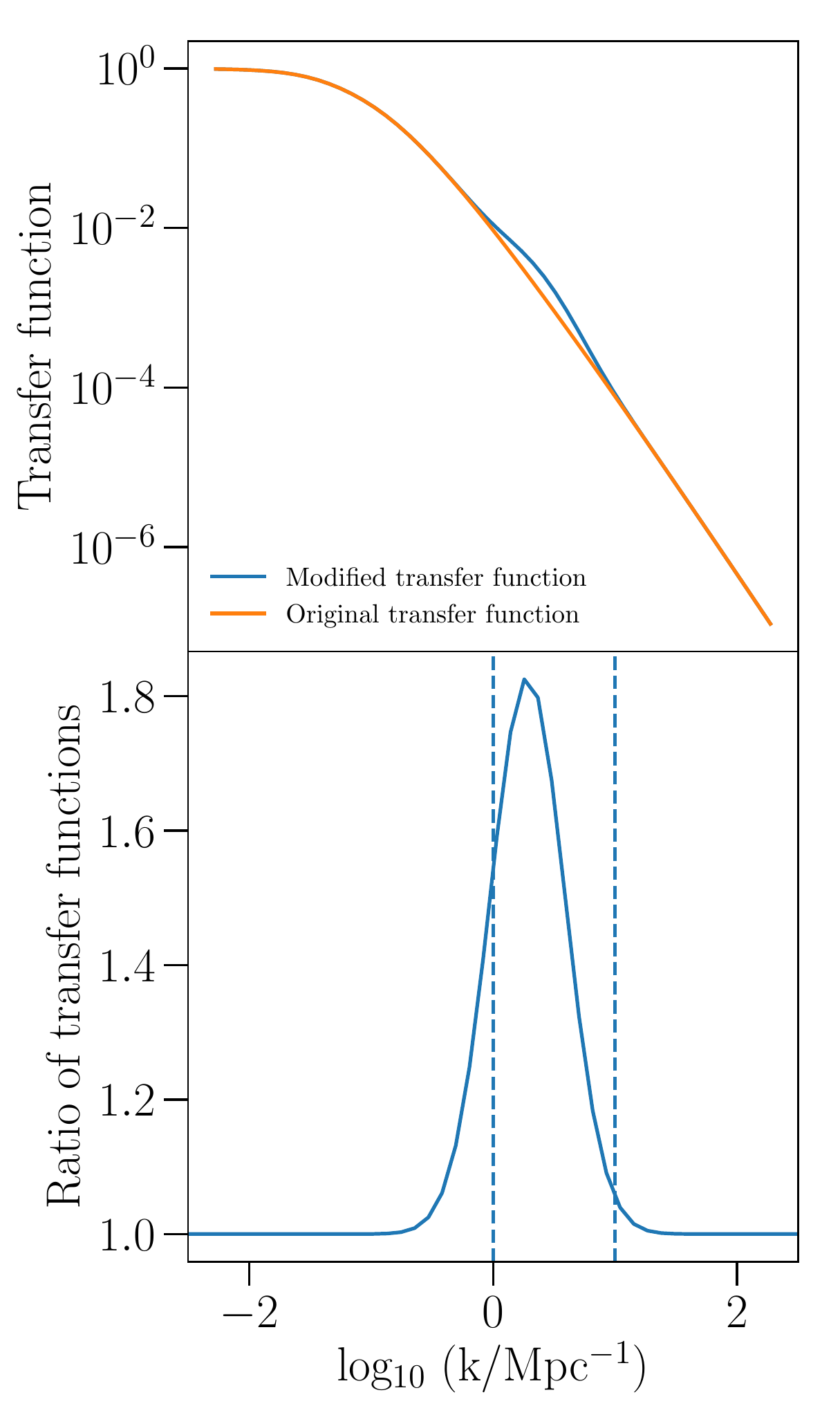}
\caption{\textit{Top panel:} Original and modified forms of the transfer function used in computing the power spectrum for halo abundances. \textit{Lower panel:} Ratio of the modified and original transfer functions in the relevant $k$-range. The region enclosed by $1 < k/({\rm Mpc} ^{-1}) < 10$ is shown by the dashed lines.}
\label{fig:transmod}
\end{figure}

\begin{figure}
\centering
\includegraphics[width = \columnwidth]{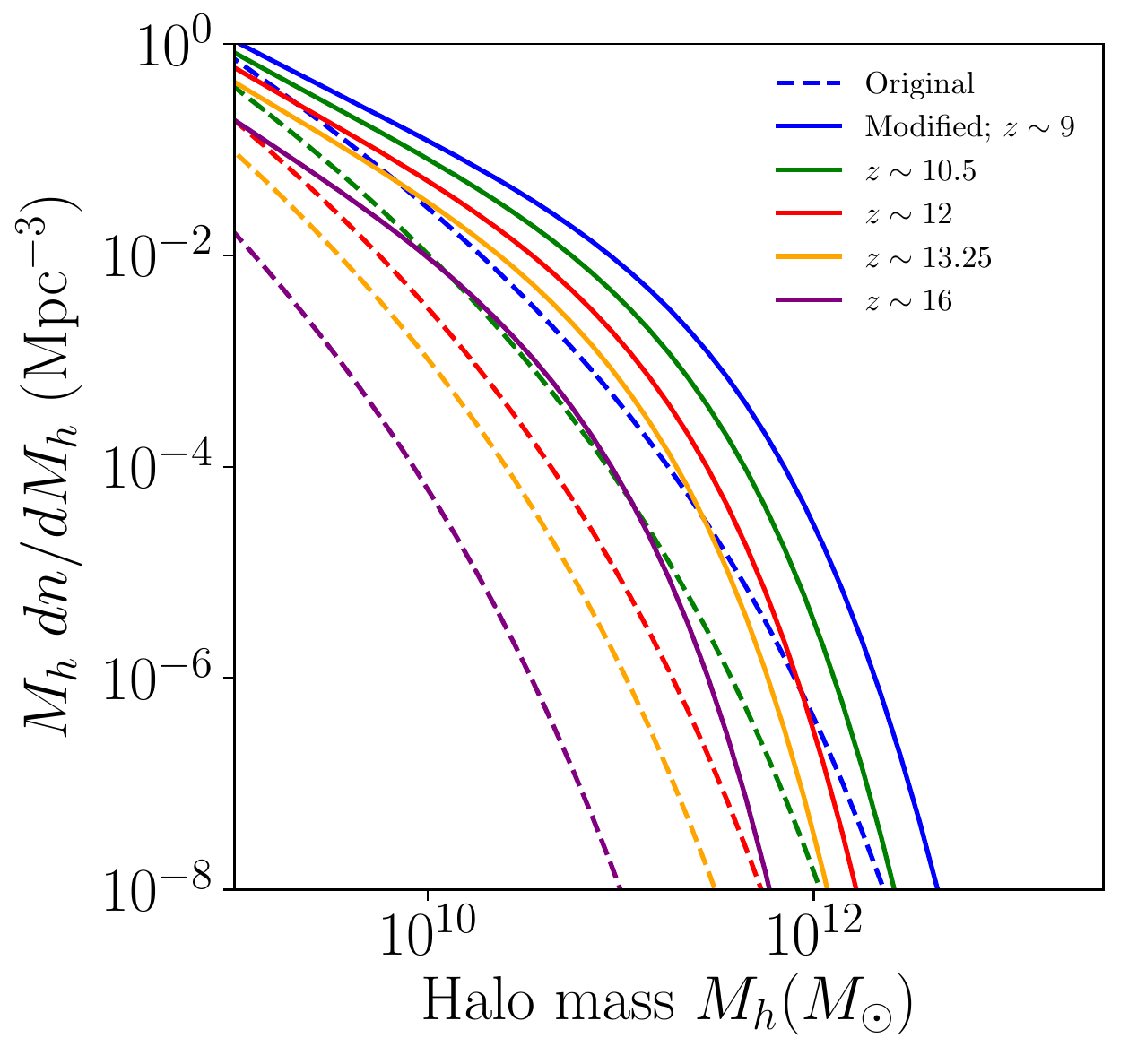}
\caption{ Modified [solid, \eq{dndmmod}] and original (dashed) halo abundances over $z \sim 9-16$ resulting from the modification to the  transfer function, \eq{transfermod}.}
\label{fig:dndmmod}
\end{figure}

\begin{figure}
\centering
\vskip-0.15in

\includegraphics[width = \columnwidth]{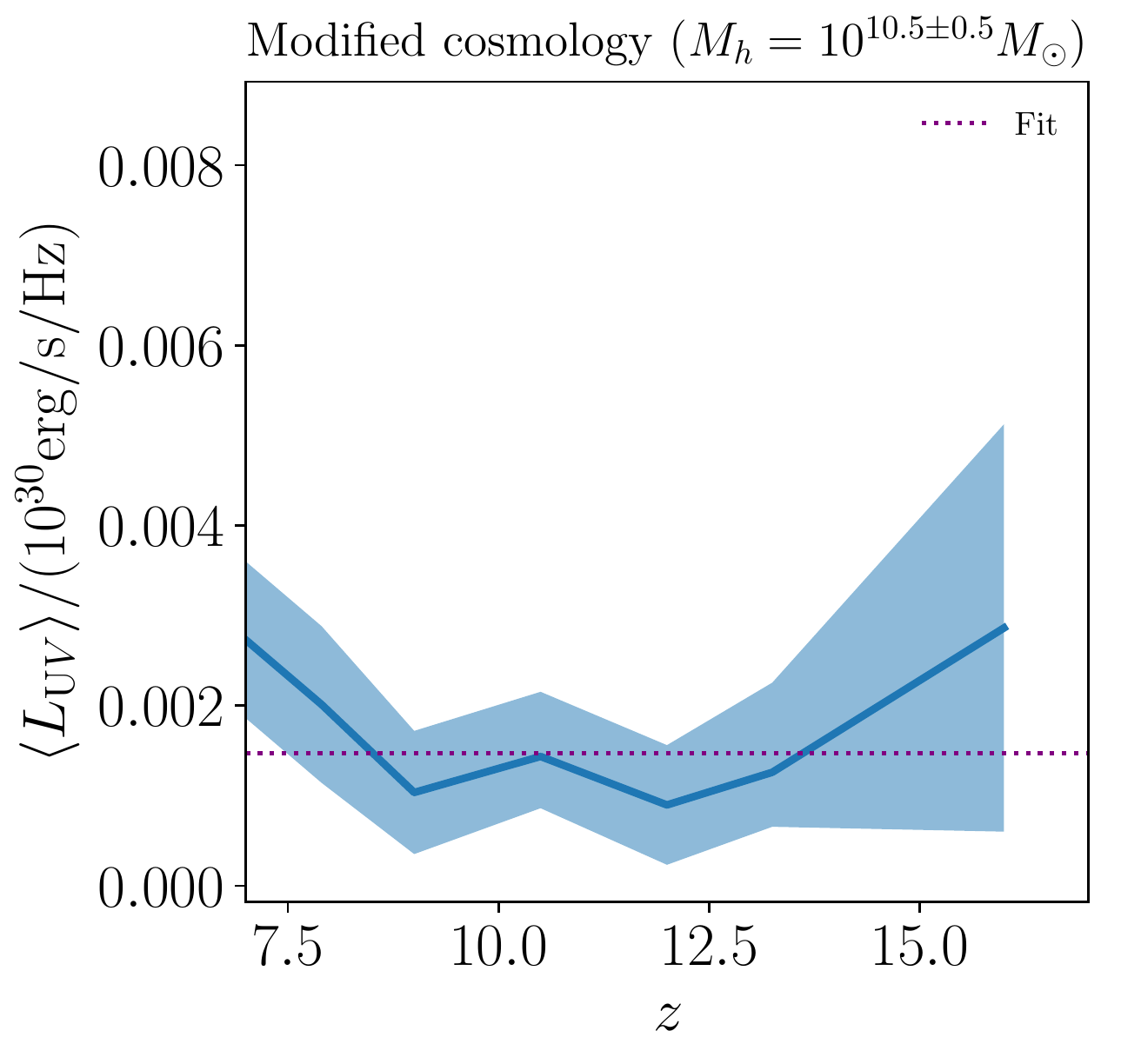}
\caption{Evolution of UV luminosity in $10^{10} - 10^{11} M_{\odot}$ haloes as a function of redshift corresponding to the modified cosmological power spectrum. The best-fitting (constant) value is shown by the dotted purple line.}
\label{fig:uvevolmod}
\end{figure}

The Pop III fraction, and that of the pristine gas over all haloes predicted by \citet{sun2021} and \citet{furlanetto2005} are shown in Fig. \ref{fig:uvevol}.  A possible proportionality to the expected gas accretion rate at fixed halo mass, following a power law $\propto (1+z)^{5/2}$ \citep{furlanetto2022} is also shown. Both of these are too gradual to explain the observed exponential rise.

Other possibilities to explain the observations include some of the objects being supernovae or hidden Active Galactic Nuclei (AGN). Most of the reported photometric spectra \citep{donnan2023, harikane2023}  do not show evidence for such behaviour, excluding the CEERS 1019 and GN-z11 spectra \citep{harikane2023b}. The NIRSpec redshifts of similar galaxies in the field \citep{curtislake2023, haro2023, fujimoto2023}  are very different from the forms expected for supernovae \citep{bufano2009}.
While eventual X-ray/transient spectra may be required to confirm the possible presence of AGN/supernovae, we can make the following predictions for the expected luminosity of an AGN in the 0.5-7 keV band (assuming the best-fitting UV magnitude inferred from \eq{uvfit} can be wholly attributed to the AGN), following a similar analysis for the double-peaked Lyman-alpha emitter COLA-1 at $z \sim 6.6$ \citep{hploeb2021}. The resultant values are quoted as lines in  Fig. \ref{fig:sedjwst} overplotted on  a fiducial QSO spectrum at $z \sim 6$, normalized to $\log_{10} (\nu L_\nu/{\rm ergs \ s}^{-1})$ (4500 \AA) = 45.5 from the observations of \citet{shen2020}. For comparison, the predicted X-ray luminosity of a COLA-1 like object \citep{hploeb2021} lies close to the $z \sim 16$ line.\footnote{We also examined the effect of a possible magnification bias on the observations. Given that the magnifications  are taken into account when calculating the luminosities in \citet{harikane2023} with most being around unity -- except for one field, SMACS J0723, whose galaxies were specifically corrected -- any possible magnification bias is expected to play a minor role in explaining the observed rise.}

\section{Modified transfer function}
\label{sec:modtrans}
Given that the astrophysical explanations are difficult to reconcile with the observed exponential rise of the UV luminosity with redshift, we now examine the possibility that the halo abundances  follow a modified $\Lambda$CDM model or -- in other words -- that the high luminosity objects are in much smaller haloes which are more abundant \citep[e.g.,][]{liu2022, sabti2023}. This approach takes advantage of the exponential tail of the halo mass function for Gaussian fluctuations, which is enhanced by an increase in the power spectrum. We consider this possibility by modifying the transfer function from its standard form, to include an enhancement around $k \sim 1 \ {\rm Mpc} ^{-1}$. The modification is assumed to follow a lognormal form:
\begin{eqnarray}
TF_{\rm mod} (k) = (1 + f(k)) TF (k);  \, \nonumber\\
f(k) =  \frac{1}{2 \sigma_{\rm tf}^2}\left(\frac{A_{\rm tf}}{\sigma_{\rm tf} k \sqrt{2\pi}}\right) \exp(-(\ln(k) - \mu)^2)
\label{transfermod}
\end{eqnarray}
with the parameters $A_{\rm tf} = 0.001, \mu = 2.303/2, \ \sigma_{\rm tf} = 2.303/50$. This is plotted in Fig. \ref{fig:transmod} with respect to the original transfer function.\footnote{Note that the maximum of this function occurs at $k \sim 1.9$ Mpc$^{-1}$, corresponding to $\ln(k) = \mu - 0.5$.} The power spectrum, $P(k)$ is calculated using the modified form of the transfer function in the usual manner:
$P_{\rm mod}(k) \propto TF_{\rm mod}(k)^2$
 and normalized to the current value of $\sigma_8$. This, is, in turn used to calculate the halo abundances via:
\begin{eqnarray}
\frac{dn}{dM_h}\Big{|} _{\rm mod} \equiv  n_{\rm mod} (M_h)  &=& A' \sqrt{\frac{2 a'}{\pi}} \frac{\rho_m}{M_h} \frac{- d \ln \sigma_{\rm mod}}{d M_h} \nu_{c, \rm mod} \nonumber \\
&& \left[1 + \frac{1}{(a' \nu_{c,\rm  mod}^2)^{q'}}\right] 
e^{-a' \nu_{c,\rm mod}^2/2} \nonumber ; \\
{\rm with} \ \nu_{\rm c, mod} = \delta_c(z)/\sigma_{\rm mod} \, ,
\label{dndmmod}
\end{eqnarray}
where $\rho_m$ is the comoving matter density and the parameters have the values $A' = 0.322$, $a' = 0.707$, and $q' = 0.3$. { The modified halo abundances are plotted as the solid lines in Fig. \ref{fig:dndmmod} in comparison to the original ones (dashed lines) at each redshift. It can be seen that the size of the modification at halo masses $10^{10} - 10^{11} M_{\odot}$ increases with increasing redshift.} 
The modified abundances are now used to infer the UV luminosity - halo mass using \eq{abmatchuvlf}. 

With the modified form of the halo abundances as above, the average UV luminosities in the $10^{10} - 10^{11} M_{\odot}$ mass range are found to be lower, and consistent with a non-evolving behaviour (as in the $z < 8$ case) across the whole redshift range as shown by Fig. \ref{fig:uvevolmod}. The best-fitting constant form is shown by the dotted line, having the value
\begin{equation}
\log_{10} (L_{\rm UV}/ {\rm ergs/s/Hz}) = 27.17 \pm 0.1
\end{equation}
{ We briefly comment on the expected impact of the above modification on the lower-$z$ luminosity functions constrained by Hubble Space Telescope (HST) observations \citep{sabti2022, sabti2022a}. As pointed out by \citet{harikane2023b}, the UV luminosity functions at $z \sim 9$ agree with those measured from HST within uncertainties, including the cosmic variance.  Also, as seen from Fig. \ref{fig:dndmmod}, the difference between the original and modified mass functions decreases with decreasing redshift, approaching about a factor of a few at $z \sim 9$ [with the characteristic luminosity at $z \lesssim 8$ being about a factor 2-3 lower than that of the original scenario, well within the uncertainties of the HST luminosity function at lower $z$ \citep{bouwens2015}.] Hence, the effect of the modification at lower redshifts are expected to be within the observational uncertainties.}

\section{Discussion} 
\label{sec:discussion}
We have illustrated the sharp rise in the UV luminosity of galaxies in haloes of masses $10^{10} - 10^{11} M_{\odot}$ over $z \sim 10-16$ inferred from the recent JWST photometric survey data. While this rise is difficult to reconcile with astrophysical expectations of the evolution of the star formation rate including the fraction of Population III stars, and the corresponding results at lower redshifts $4 < z < 8$ \citep{mashian2015} that indicate a fairly non-evolving UV luminosity and star formation rate as a function of halo mass, we find that an enhancement in the power by a factor of order a few on scales corresponding to $k \sim 1$ Mpc$^{-1}$ -- encapsulated by a modified $\Lambda$CDM power spectrum -- is sufficient to reproduce a non-evolving form of the UV luminosity functions across all redshifts.  This explanation predicts that the exponential deviation from a constant $L_{\rm UV} (M_h)$ will still continue in $\Lambda$CDM at redshifts higher than $z \sim 16$. 

The inferred enhancement could be of astrophysical or cosmological origin. The comoving mass scale -- which may be considered as an effective `filtering' or Jeans' mass  associated with the peak $k$-value, $k \sim 1.9 \ {\rm Mpc}^{-1}$ -- needed to alleviate the rise is { $M_{\rm fs} = ({4 \pi }/{3})\rho_m (\pi/ k)^3 \sim 6.7 \times 10^{11} M_{\odot}$} where $\rho_m$ is the current mean dark matter density. Noting that the enhancement required is only by a factor of a few, it is also consistent with astrophysical effects of radiative feedback on the clustering of high-luminosity haloes \citep[e.g.,][]{sabti2023, chen2023}. Future spectroscopic data from the JWST surveys will { help improve  the statistics  needed  to verify the redshift evolution in the UV luminosity functions at $z > 10$, as well as validate or rule out astrophysical explanations involving the presence of hidden AGN and/or Pop III candidates \citep{harikane2023,harikane2023c}.}

\section*{Acknowledgements}  
 
 HP's research is supported by the Swiss National Science Foundation via Ambizione Grant PZ00P2\_179934. The work of AL is supported in part by the Black Hole Initiative, which is funded by grants from the John Templeton Foundation and the Gordon and Betty Moore Foundation. We thank the anonymous referee for a helpful report. 

\def\aj{AJ}                   
\def\araa{ARA\&A}             
\def\apj{ApJ}                 
\def\apjl{ApJ}                
\def\apjs{ApJS}               
\def\ao{Appl.Optics}          
\def\apss{Ap\&SS}             
\def\aap{A\&A}                
\def\aapr{A\&A~Rev.}          
\def\aaps{A\&AS}              
\def\azh{AZh}                 
\def\baas{BAAS}
\def\jcap{JCAP}
\def\jrasc{JRASC}             
\def\memras{MmRAS}
\def\na{New Astronomy}
\def\nat{Nature}
\def\mnras{MNRAS}             
\def\pra{Phys.Rev.A}          
\def\prb{Phys.Rev.B}          
\def\prc{Phys.Rev.C}          
\def\prd{Phys.Rev.D}          
\def\prl{Phys.Rev.Lett}       
\def\pasp{PASP}               
\def\pasj{PASJ}
\def\physrep{Phys. Repts.}
\def\qjras{QJRAS}             
\def\skytel{S\&T}             
\def\solphys{Solar~Phys.}     
\def\sovast{Soviet~Ast.}      
\def\ssr{Space~Sci.Rev.}      
\def\zap{ZAp}                 
\let\astap=\aap
\let\apjlett=\apjl
\let\apjsupp=\apjs

\small{
\bibliographystyle{plainnat}
\bibliography{mybib}
}

\end{document}